\documentclass[aps,prl,showpacs,groupedaddress,twocolumn,amsmath,amssymb]{revtex4}

\usepackage[american]{babel}
\usepackage{graphicx}
\usepackage{times}

\newcommand{\ud}{\mathrm{d}}

\begin{document}
\title{Experimental determination of a nonclassical Glauber-Sudarshan P function}

\author{T. Kiesel and W. Vogel}

\affiliation{Arbeitsgruppe Quantenoptik, Institut f\"ur Physik, Universit\"at  Rostock, D-18051 Rostock,
Germany}

\author{V. Parigi}
\affiliation{Department of Physics, University of Florence, I-50019 Sesto Fiorentino, Florence,Italy}
\affiliation{LENS, Via Nello Carrara 1, 50019 Sesto Fiorentino, Florence, Italy}

\author{A. Zavatta$^{2,3}$ and M. Bellini$^{2,3}$}
\affiliation{ LENS, Via Nello Carrara 1, 50019 Sesto Fiorentino, Florence, Italy}
\affiliation{Istituto Nazionale di Ottica Applicata, CNR, L.go E. Fermi, 6, I-50125, Florence, Italy}

\begin{abstract}
A quantum state is nonclassical if its Glauber-Sudarshan $P$ function fails to be interpreted as a
probability density. This quantity is often highly singular, so that its reconstruction is a demanding
task. Here we present the experimental determination of a well-behaved $P$ function showing
negativities for a single-photon-added thermal state. This is a direct visualization of the original
definition of nonclassicality. The method can be useful under conditions for which many other signatures of nonclassicality would not persist.
\end{abstract}


\maketitle

Einstein's hypothetical introduction of light quanta, the photons, was the first step toward the
consideration of nonclassical properties of radiation~\cite{annphys17-132}. But what does nonclassicality mean in a general sense? A radiation field is called nonclassical when its
properties cannot be understood within the framework of the classical stochastic theory of
electromagnetism. For other systems, nonclassicality can be defined accordingly. Here we will focus our
attention on harmonic quantum systems, such as radiation fields or quantum-mechanical oscillators, for
example trapped atoms. 

In this context the coherent states, first considered by Schr\"odinger in the
form of wave packets~\cite{natwi14-664}, play an important role.  They represent those quantum states
that are most closely related to the classical behavior of an oscillator or an electromagnetic wave. For
a single radiation mode, the coherent states $|\alpha\rangle$ are defined as the right-hand eigenstates
of the non-Hermitian photon annihilation operator $\hat{a}$, $\hat{a} |\alpha \rangle = \alpha |\alpha
\rangle$; cf. e.g.~\cite{pr131-2766}. A general mixed quantum state $\hat{\rho}$,
\begin{equation}\label{eq:P}
\hat{\rho} = \int d^2 \alpha\, P(\alpha)  |\alpha \rangle \langle \alpha |,
\end{equation}
can be characterized by the Glauber-Sudarshan $P$ function~\cite{pr131-2766,prl-10-277}. In this form
the quantum statistical averages of normally ordered operator functions can be written as
\begin{equation}\label{eq:av}
\langle : \hat{f} (\hat{a},\hat{a}^\dagger ) : \rangle = \int d^2 \alpha P(\alpha) f (\alpha,
\alpha^\ast),
\end{equation}
where the normal ordering prescription $: \hat{f} (\hat{a},\hat{a}^\dagger) :$ means that all creation
operators $\hat{a}^\dagger$ are to be ordered to the left of all annihilation operators $\hat{a}$.

Formally, the resulting expressions~(\ref{eq:av}) for expectation values are equivalent to classical
statistical mean values. However, in general, the $P$ function does not exhibit all the properties of a
classical probability density. It can become negative or even highly singular. Within the chosen
representation of the theory, the failure of the Glauber-Sudarshan $P$ function to show the properties
of a probability density is taken as the key signature of quantumness~\cite{pr-140B-676,scr-T12-34}.

In this Rapid Communication we demonstrate the experimental determination of a nonclassical $P$ function. Within the experimental precision it clearly attains negative values. This is a direct demonstration of
nonclassicality: the negativity of the $P$ function prevents its interpretation as a classical
probability density.

Why is it so difficult to demonstrate the nonclassicality directly on the basis of this original
definition? Let us go back to a single photon as postulated by Einstein. Its $P$ function
is 
\begin{equation}\label{eq:P-1ph}
  P(\alpha) = \left ( 1 +
  \frac{\partial}{\partial\alpha} \,
  \frac{\partial}{\partial \alpha^{\ast}} \right )\, \delta(\alpha);
\end{equation}
cf. e.g.~\cite{vowe-book}. Already in this case we get a highly singular distribution in terms of
derivatives of the $\delta$ distribution, which cannot be interpreted as a classical probability. Due to these properties, it is difficult to experimentally determine nonclassical $P$ functions in general.

How can one realize nonclassical states whose properties can be demonstrated directly in terms of the
original definition, that the $P$ function fails to be a probability density? 
This question is not trivial: for instance, losses introduced by imperfect experimental efficiencies lead only to rescaling of the quadrature variable; cf., e.g., \cite{jopB39-905}. The $P$ function obtained by perfect detection
is related to $P_\eta(\alpha)$, obtained with the quantum efficiency $\eta$ via 
\begin{equation}
	P(\alpha) = \eta P_\eta(\sqrt{\eta}\alpha).\label{eq:P:and:losses}
\end{equation}
Consequently, singularities in the $P$ function are then preserved. Most of the nonclassical states experimentally generated so far have highly singular $P$ functions, whose reconstruction is impossible. 
However, one may start with a thermal state $\hat{\rho}_{\rm th}$ with mean
photon number $\bar n$. By photon creation one gets a single-photon-added thermal state (SPATS), $\hat{\rho} = {\cal N} \hat{a}^\dagger
\hat{\rho}_{\rm th} \hat{a}$, where $\cal{N}$ denotes the normalization. Its $P$ function is now
well behaved, but violates the properties of a classical probability density~\cite{pra-46-485},
\begin{equation}\label{eq:P-spats}
  P(\alpha) = \frac{1}{\pi\bar n^3}\left[(1+\bar n)|\alpha|^2-\bar n\right]e^{-|\alpha|^2/\bar n},
\end{equation}
giving rise to the question whether its experimental determination could be possible~\cite{jmo-48-1881}.
In the zero-temperature limit, the SPATS includes the special case of the one-photon Fock state with the
highly singular $P$ function given in Eq.~(\ref{eq:P-1ph}). In this sense the SPATS represents a single
photon whose $P$ function is regularized by a controlled thermal background. 

Recently, SPATSs could be realized experimentally and some of their nonclassical signatures have been
verified~\cite{pra75-052106}. Nevertheless, the reconstruction of a nonclassical $P$ function remains a
challenging problem which goes beyond the standard procedures of quantum state reconstruction, for the
latter. see, e.g.,~\cite{vowe-book}. A successful determination of the $P$ function of a SPATS would
visualize the basic definition of nonclassicality for a quantum state that lies at the heart of
Einstein's hypothesis: a regularized version of a single photon.

\begin{figure}[h]
\includegraphics*[width=\columnwidth]{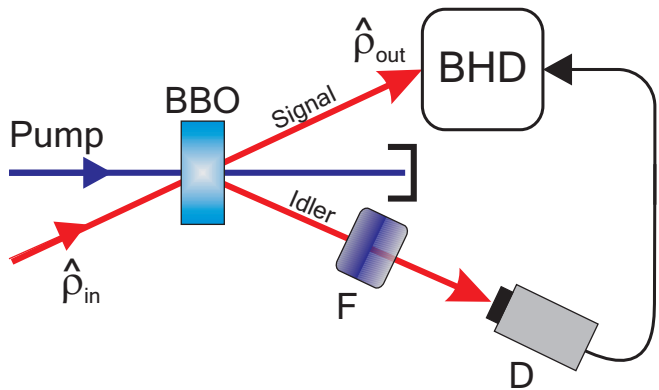}
\caption{Scheme for the conditional excitation of a thermal light state (denoted by $\hat \rho_{\rm in}$) by
a single photon. A click in the on-off detector D prepares the photon-added thermal state $\hat
\rho_{\rm out}$ and triggers its balanced homodyne 
detection (BHD). \label{fig:setup}}
\end{figure}

The core of the experimental apparatus used to produce SPATSs is an optical parametric amplifier based
on a type-I $\beta$-barium borate (BBO) crystal pumped by radiation at 393 nm (see
Fig.~\ref{fig:setup}). The pump is obtained by second harmonic generation in a lithium triborate (LBO) crystal of a
mode-locked Ti:sapphire laser emitting 1.5 ps pulses with a repetition rate of 82 MHz. When the parametric
amplifier is not injected, spontaneous parametric down-conversion takes place, generating pairs of
photons at the same wavelength as the laser source along two directions commonly called the signal and idler
channels. We perform a conditional preparation of the quantum states by placing an on-off photodetector
(D) after narrow spectral-spatial filters (F) along the idler channel~\cite{pra04,pra75-052106}.
A click of the idler detector prepares the signal state, whose quadratures are measured on a pulse-to-pulse basis using an ultrafast balanced
homodyne detection scheme~\cite{josab02}. After verifying the phase independence of the quadrature distributions, the state is then analyzed by acquiring quadrature values with random local oscillator phases.
When no fields are present at the inputs of the parametric amplifier, conditioned single-photon Fock
states are spontaneously generated in the signal channel~\cite{lvovsky01,pra04}. On the other hand, we
have recently shown that the injection of pure or mixed states results in the conditional production of
their single-photon-added versions, always converting the initial states into nonclassical ones~\cite{science04,pra75-052106,science07}.

Here we use a pseudo thermal source, obtained by inserting a rotating ground glass disk in a
portion of the laser beam, for injecting the parametric amplifier and producing SPATSs. The scattered
light forms a random spatial distribution of speckles whose average size is larger than the core
diameter of a single-mode fiber used to collect it. When the ground glass disk rotates, light exits the
fiber in a clean collimated spatial mode with random amplitude and phase fluctuations, yielding the
photon distribution typical of a thermal source~\cite{arecchi65}. The product between the SPATS
preparation rate and the co\-he\-ren\-ce time of the injected thermal state (a few microseconds, and depending
on the rotation speed of the disk) is kept much smaller than 1. This condition assures that each state
is prepared by adding a single photon to a coherent state having an amplitude and phase which are
completely uncorrelated with respect to those of the previous one. This experimental realization of a
thermal state directly recalls its P function definition, i.e., a statistical mixture of coherent states
weighted by a Gaussian distribution: $P(\alpha)=\exp(-|\alpha|^2/\bar{n})/(\bar{n}\pi)$.

By performing measurements on single-photon Fock states and on unconditioned thermal ones, we have
estimated an overall experimental efficiency of 0.62$\pm 0.04$. Both the li\-mi\-ted efficiency in the state
preparation ($\approx$0.92) and in homodyne detection ($\approx$0.67) degrade the expected final state
by introducing unwanted losses. This does not contaminate the obtained
$P$ function; cf. Eq.~(\ref{eq:P:and:losses}).


Let us now proceed with the reconstruction of the $P$ function. 
Its characteristic function $\Phi(\beta)$ is related to that of the quadrature $\hat{x}(\varphi)$~\cite{vowe-book},
\begin{equation}
        \Phi(\beta) = \left<:\hat D(\beta):\right> = \left<e^{i|\beta|\hat x(\pi/2-\arg(\beta))}\right> e^{|\beta|^2/2},
        \label{eq:charfunc:P}
\end{equation}
where $\hat D(\beta)$ is the displacement operator. 
Since the measured state is independent of phase, we may neg\-lect the arguments of $\beta$ and $\hat{x}$. The expectation value on the right-hand side represents the characteristic function of the observable quadrature. It can be estimated from the sample of $N$ measured quadrature values $\left\{x_j\right\}_{j=1}^N$
via (cf.~\cite{lvovskyShapiro})
\begin{equation}
        \left<e^{i|\beta|\hat x}\right> \approx \frac{1}{N}\sum_{j=1}^{N}e^{i|\beta|x_j}.
        \label{eq:charfunc:x}
\end{equation}
Inserting Eq.~(\ref{eq:charfunc:x}) into~(\ref{eq:charfunc:P}), we get an estimation $\overline{\Phi}(\beta)$ of ${\Phi}(\beta)$.
The variance of this quantity can be estimated as 
\begin{equation}
        \sigma^2\left\{\overline{\Phi}(\beta)\right\}  = \frac{1}{N}\big[e^{|\beta|^2}-\left|\overline{\Phi}(\beta)\right|^2\big].
\end{equation}

The inverse Fourier transform of $\Phi(\beta)$ yields the $P$ function, which for many nonclassical states does not exist as a well-behaved function.
However, the sampled characteristic function converges stochastically toward the theoretical one. In our case its Fourier transform is an ana\-ly\-ti\-cal function. For radial symmetry of the state the 
two-dimensional Fourier transform reduces to the Hankel transform~\cite{Jerri},
\begin{equation}
        P(\alpha)  = \frac{2}{\pi}\int_0^\infty b J_0(2b|\alpha|) \Phi(b)\ud b. \label{eq:Phi:to:P:Hankel}
\end{equation}
In our treatment we set the experimental curve to zero for arguments greater than a cutoff value $|\beta|_c$, where the graph becomes small. This limits the disturbing sampling noise
on the reconstructed function
\begin{equation}
        \overline{P}(\alpha)  = \frac{2}{\pi}\int_0^{|\beta|_c} b J_0(2b|\alpha|) \overline{\Phi}(b)d b \label{eq:Phi:to:P:Hankel:exp}
\end{equation}
 to a reasonable level. 
The corresponding variance has been calculated as
\begin{widetext}
        \begin{equation}
        \sigma^2\left\{\overline{P}(\alpha)\right\} = 
                \frac{1}{N}\left(\frac{4}{\pi^2}\iint_0^{|\beta|_c}\!\!\!\! b b' J_0(2b|\alpha|)J_0(2b'|\alpha|) \overline{\Phi}(b-b')e^{b b'}d b\,d b' - \overline{P}(\alpha)^2\right).
        \label{eq:app:var:P:Hankel}
        \end{equation}
\end{widetext}
The systematic error
\begin{equation}
        \Delta_P(\alpha)  = \frac{2}{\pi}\int_{|\beta|_c}^\infty b J_0(2b|\alpha|) \Phi(b) d b \label{eq:Phi:to:P:Hankel:syst:err}
\end{equation}
is estimated with the help of the fitted theoretical function. 

\begin{figure}[h]
        \includegraphics[width=\columnwidth]{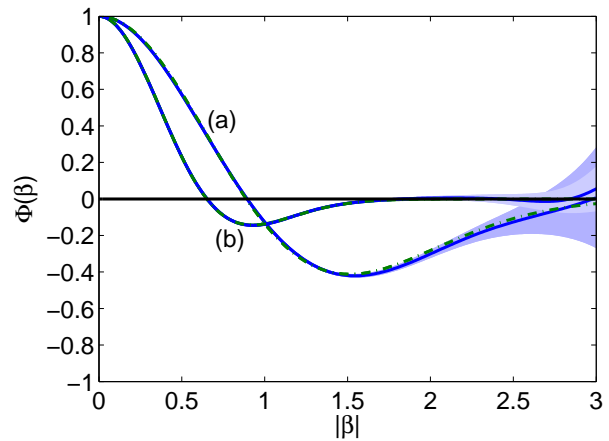}
        \caption{Experimental characteristic functions (solid lines) and best fit to theoretical curves (dashed lines): (a) SPATS , with $\bar n=1.11$ and $\eta=0.60$, (b) mixture of SPATS with 19\% of the thermal background, with $\bar n=3.71$ and $\eta=0.62$. The shaded areas show the standard deviations.}
        \label{fig:charfunc}
\end{figure}

In Fig.~\ref{fig:charfunc} we show experimental curves for characteristic functions. Curve (a) is in good agreement with the expected characteristic function $\Phi(\beta)$ for a SPATS,
\begin{equation}
        \Phi(\beta) = \left[1-(1+\bar n)|\beta|^2\right]e^{-\bar n|\beta|^2},       \label{eq:CF-spats}
\end{equation}
for the mean thermal photon number  $\bar n = 1.11$ and the global quantum efficiency $\eta=0.60$. Curve (b) shows the cha\-rac\-te\-ris\-tic function for a mixture of a SPATS and its thermal background with weights of 0.81 and 0.19, respectively, for 
$\bar n = 3.71$ and $\eta=0.62$. For sampling these functions, we have acquired $10^5$ and $5\times 10^5$ data points for the curves (a) and (b) respectively.
We note that both curves are suited to reconstruct the corresponding $P$ functions by properly choosing cutoff values $|\beta|_c$ of their arguments.

\begin{figure}[h]
        \includegraphics[width=\columnwidth]{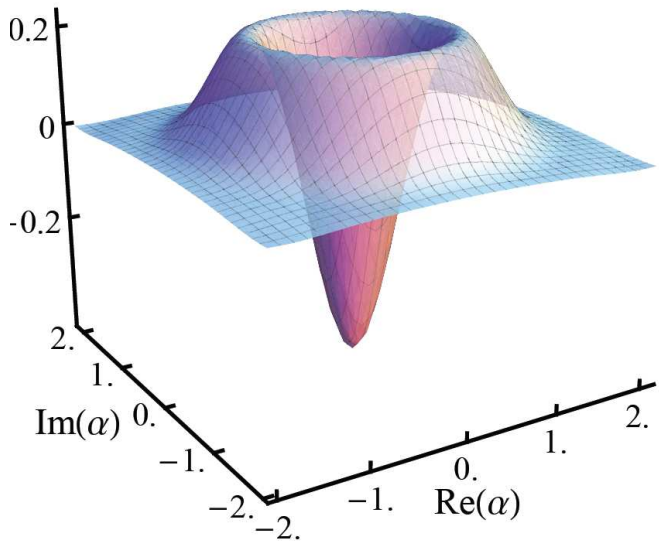}
        \caption{Experimentally reconstructed $P$ function of a SPATS, as obtained from Fig.~\ref{fig:charfunc}(a).}
        \label{fig:P:3D}
\end{figure}

The reconstructed $P$ function, shown in Fig.~\ref{fig:P:3D}, is derived from the experimental
characteristic function given in Fig.~\ref{fig:charfunc}(a). Since the measured states are independent of the phase, the
reconstructed $P$-representation is phase independent as well. It is clearly seen that the $P$ function attains
negative values, so that it fails to have the properties of a classical probability density. This is a direct proof of the nonclassicality of the experimentally realized SPATS, based on the original definition of nonclassicality~\cite{pr-140B-676,scr-T12-34}.

\begin{figure}[h]
        \includegraphics[width=\columnwidth]{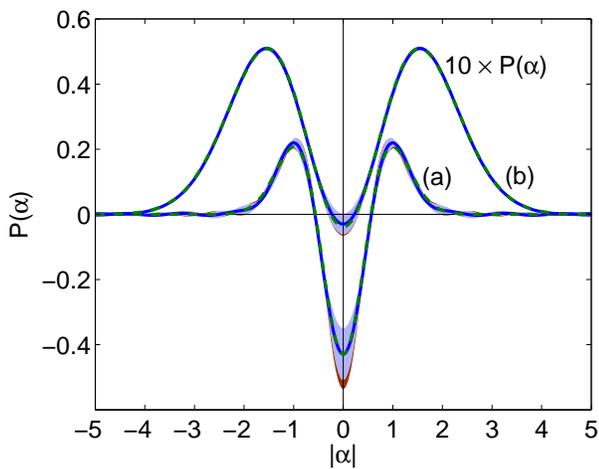}
        \caption{The $P$ functions (solid lines) in parts (a) and (b) are obtained from the experimental characteristic functions in Figs.~\ref{fig:charfunc}(a) and \ref{fig:charfunc}(b) respectively. They are compared with the corresponding theoretical fits (dashed curves). The standard deviations (light shaded areas) and the systematic errors (dark shaded areas) are also given.}
        \label{fig:P:crosssection}
\end{figure}

For a more careful discussion, we also examine a cross section along a radial line, as shown in
Fig.~\ref{fig:P:crosssection}(a). The experimentally determined curve is drawn with the solid line.
Obviously, it is in good agreement with the theoretical expectation (dashed curve). The distance between the minimum value and the $|\alpha|$ axis is approximately equal to five standard deviations, which is not diminished by the systematic error of $|\Delta_P(\alpha)| < 0.07 |P(0)|$, obtained by the cutoff $|\beta|_c=2.8$.
The statistically significant negativity of the $P$ function prevents it from being interpreted as a classical probability density. This provides a clear evidence of nonclassicality per definition.

Special nonclassical signatures of SPATSs, which are consequences of the negativities of the $P$ function, have been experimentally demonstrated recently~\cite{pra75-052106}. It is important to note that the reconstruction of the $P$ function is just possible for sufficiently large thermal photon number $\bar{n}$. On the contrary, other criteria for nonclassicality, such as negativities of the Wigner function, the Klyshko criterion, and the entanglement potential, start to fail for increasing values of $\bar{n}$. To show the power of the reconstruction of the $P$ function under such conditions, we have demonstrated its use at the limits: for a SPATS with $\bar{n}=3.71$, which is additionally contaminated with a 19\% admixture of the corresponding thermal background. By using a cutoff $|\beta|_c = 1.9$, we still obtain a $P$ function being negative within one standard deviation. Other nonclassical effects, as discussed above, do not survive for this state.

Criteria for the characteristic functions are known, which are equivalent to the negativity of the $P$ function \cite{prl-89-283601}. For many states 
the characteristic function displays their nonclassicality by violating the condition
$|\Phi(\beta)| \le 1$, cf.~\cite{prl-84-1849}. If the condition is satisfied, $\Phi(\beta)$ may be integrable and then the $P$ function can be obtained to directly verify nonclassicality. SPATSs belong to this category: for sufficiently high $\bar n$ most criteria for nonclassicality (including the lowest-order one based on the characteristic function) fail~\cite{pra75-052106}, but it is still possible to retrieve a negative $P$ function.

Let us consider how sensitively the negativities of the $P$ function depend on the overall efficiency $\eta$. Ba\-lanced homodyne detection measures the ``true'' state quadratures when the efficiency
is unity. For imperfect detection ($\eta <1$) one records a convolution of the
quadrature distribution with Gaussian noise, whose variance increases with decreasing $\eta$; cf.~\cite{pra-47-4227}. In the Wigner function, this
increasing noise smoothes out its structures and may destroy their negativities. As can be seen from Eq.~(\ref{eq:P:and:losses}), the shape and
the relative noise level of the reconstructed $P$ function do not depend on the efficiency. Hence the negativities of $P(\alpha)$
are in principle preserved even for a small efficiency, whereas for other phase-space distributions, such as the Wigner function, they are quickly lost.

In conclusion, we have reconstructed the Glauber-Sudarshan $P$ function of an experimentally prepared
single-photon-added thermal state. We obtain a well-behaved function with statistically significant
negativities, so that it fails to show the properties of a classical probability density. This is a
direct demonstration of nonclassicality according to its original definition. The approach works well, just when many other methods of demonstrating nonclassicality fail.

This work was partially supported by Ente Cassa di Risparmio di Firenze and CNR, under the RSTL
initiative.

\end{document}